\documentclass[
  journal=largetwo,
  manuscript=Research-Article,
  year=2024,
  volume=xx,
]{cup-journal}

\usepackage{amsmath}
\usepackage[nopatch]{microtype}
\usepackage{booktabs}
\usepackage[T1]{fontenc}
\usepackage{graphicx}	

\newcommand{\romn}[1] {{\mathrm #1}}
\newcommand\fm{\hbox{$.\!\!^{\reset@font\romn m}$}}

\newcommand\fd{\hbox{$.\!\!^{\rm d}$}}

\title{The multiple nature of CC Com: one of the ultra-short orbital period late-type contact binary systems}

\author{Dolunay Ko\c{c}ak}
\affiliation{Institute of Astronomy, University of Cambridge, Madingley Road, Cambridge CB3 OHA, UK}
\alsoaffiliation{Department of Astronomy and Space Sciences, Faculty of Science, Ege University, 35100, {\.I}zmir, T\"urkiye}
\email[Dolunay Ko\c{c}ak]{dkocak@ast.cam.ac.uk}

\DeclareBibliographyDriver{software}{...}


\begin{document}

\begin{abstract}
The study of very short-period contact binaries provides an important laboratory in which the most important and problematic astrophysical processes of stellar evolution take place. Short-period contact systems, such as CC Com are particularly important for binary evolution. Close binary systems, especially those with multiple system members, have significant period variations, angular momentum loss mechanisms predominance, and pre-merger stellar evolution, making them valuable astrophysical laboratories. In this study, observations of CC Com, previously reported as a binary system, and new observations from the T\"UB\.ITAK National Observatory (TUG) and the space-based telescope TESS have revealed that there is a third object with a period of about eight years and a fourth object with a period of about a century orbiting the binary system. From simultaneous analysis of all available light curves and radial velocities, the sensitive orbital and physical parameters of the system components are derived. The orbital parameters of the components are P$_{\rm  A}=0.221\pm0$\ days, P$_{\rm  B}=7.9\pm0.1$\ yr, P$_{\rm  C}=98\pm5$\ yr, $e_3$ = 0.06, $e_4$ = 0.44 and the physical parameters as M$_{\rm  A1}=0.712\pm0.009$\,M$_{\odot}$, M$_{\rm  A2}=0.372\pm0.005$\,M$_{\odot}$, $m_{B;i'=90^\circ}$=0.074 M$_{\odot}$, $m_{C;i'=90^\circ}$=0.18 M$_{\odot}$, R$_{\rm  A1}=0.693\pm0.006$\,R$_{\odot}$,  R$_{\rm  A2}=0.514\pm0.005$\,R$_{\odot}$, L$_{\rm A1}$ = 0.103 L$_\odot$, L$_{\rm A2}$ = 0.081 L$_\odot$. Finally, the evolutionary status of the multiple system CC Com and its component stars is discussed. 
\end{abstract}

\section{INTRODUCTION} \label{sec:int}
Contact binaries provide valuable insights into the evolution of binary star systems. They are a stage in the evolution of close binary systems, and studying their properties helps us understand how stars evolve when they interact closely with each other \citep{Yakut2005ApJ...629.1055Y}. Contact systems are essential objects for studying the formation mechanisms of merging phenomena, very close binaries, and astrophysical processes of binary stars, as seen for V1309 Scorpii \citep{Tylenda2011A&A...528A.114T}. The close interactions between stars in contact binaries allow us to study in detail the complex dynamics that can lead to the formation of these systems and the role of mass loss and mass transfer in their evolution. In addition, extremely short-period contact binaries, such as CC Com, provide a crucial astrophysical laboratory to test the importance of angular momentum loss mechanisms such as magnetised stellar wind and gravitational radiation in the evolutionary processes of these systems \citep{Yakut2005ApJ...629.1055Y}. 
The time required for a system such as CC Com, which is composed of very small masses of 0.72 M$_{\odot}$ and 0.38 M$_{\odot}$ \citep{Kose2011AN....332..626K}, to evolve or fill the in Roche lobes through the \textit{normal} process of evolution, is significantly greater than the age of the Universe. CC Com is an excellent example because the proximity effect dominates its evolution more than other processes. Binary systems characterised by an extremely short orbital period have the potential to generate detectable gravitational waves \citep{Ju2000RPPh...63.1317J, Kose2011AN....332..626K}.

One of the more basic pieces of information we get from a star is the variation in its atmospheric layers. Late-type stars are mostly active owing to their convective layers, and they lose mass over time. This mass loss has an impact on the evolution of stars. The mass loss becomes even more important when the orbital period of the binary system is relatively short. This mass loss causes a decrease in the total mass and, so, a change in the orbital period of the system. In the later stages of binary evolution, the Roche lobe of the primary star first fills its Roche lobe and begins to transfer mass from the L1 point. With mass transfer, the mass ratio changes and this causes a change in the period. Both mass loss and mass transfer play a role as in angular momentum loss from the system. Another effect is the presence of a third or more bodies, which cause a loss of angular momentum from the binary system. This effect is explained by the von Zeipel-Lidov-Kozai effect \citep{vonZeipel1910AN....183..345V, Kozai1962AJ.....67..591K, Lidov1962P&SS....9..719L}.

CC Com has been observed with photometric and spectral methods for over half a century. CC Com was first discovered as a variable star by \citet{Hoffmeister1964AN....288...49H}.
The basic parameters of the system are given in Table \ref{Table:cccom:basic:par}. The B-V colour of the K4-5V \citep{Pribulla2007AJ....133.1977P} spectral system was calculated to be 0.54 mag \citep{Kreiner2001aocd.book.....K}. In a more detailed study of the system, \citet{Rucinski1969AcA....19..245R} obtained light curves in U, B, and V filters. Under the assumption of a being a contact system, \citet{Rucinski1969AcA....19..245R} found a light curve solution of the system. Subsequently, \citet{Breinhorst1982Ap&SS..86..107B} obtained the light curves of the system in the B and V bands. \citet{Bradstreet1985ApJS...58..413B} obtained the parameters of the system by the light curve and the solution of the radial velocity.  \citet{Zhou1988Ap&SS.141..199Z} and \citet{Linnell1989ApJ...343..909L} produced synthetic models of the system using a differential correction method. Later, the light curve solutions were updated with modern methods \citep{Zhu2021RAA....21...84Z,Kose2011AN....332..626K,Zola2010MNRAS.408..464Z,Rucinski2008MNRAS.388.1831R}, with radial velocity data from \citet{Pribulla2007AJ....133.1977P}, \citet{McLean1983MNRAS.203....1M} and \citet{Rucinski1977PASP...89..684R}. As a result of the analyses performed, \citet{Kose2011AN....332..626K} found that the masses and radii of the hot and cold components of the system were 0.378 M$_{\odot}$, 0.717 M$_{\odot}$, 0.530 R$_{\odot}$, 0.708 R$_{\odot}$, while \citet{Zhu2021RAA....21...84Z} obtained these as 0.409 M$_{\odot}$, 0.748 M$_{\odot}$, 0.550 R$_{\odot}$, 0.720 R$_{\odot}$.

The presence of asymmetry at maximum light levels was noted in many of the light curves obtained for CC Com \citep{Kose2011AN....332..626K,Kocak2023PhDT.........3K}.
 Similar variations owing to stellar spots have also been detected at minimum light. Most contact binaries show the O'Connell effect in their light curves, caused by stellar spots \citep{Yakut2005ApJ...629.1055Y}. The solution of the light curve found by \citet{Kose2011AN....332..626K} showed that 6\% of the surface of the primary star is covered by a cold spot. There are physical processes that cause changes in the orbit of a binary system, such as mass transfer, mass loss, the presence of a third body, axis rotation, stellar activity, and relativistic effects \citep{Eggleton2006epbm.book.....E,Eggleton2017MNRAS.468.3533E,Sarkar2024arXiv240205912S,Barack2019CQGra..36n3001B,Coughlin2008AJ....136.1089C,Christopoulou2011AJ....142...99C,Qian2013AJ....145...91Q,Li2019RAA....19..147L}.

The analysis of the period variation of the system under the assumptions of a limited number of time minima has been previously considered by many authors \citep{Qian2001MNRAS.328..914Q,Yang2003PASP..115..748Y,Kose2011AN....332..626K,Yang2009AJ....137..236Y,Zhu2021RAA....21...84Z}. 
The parabolic variation obtained from the analysis of the \(O-C\) diagram of the system reveals mass transfer from the massive star to the low-mass star at $1.6\times10^{-8}$ M$_{\odot}$ per year by \citet{Kose2011AN....332..626K}.  \citet{Zhu2021RAA....21...84Z} calculated that the orbital period of the system decreases by $4.66\pm0.20\times 10^{-11}$ days per year. It has been discussed that the oscillation of the orbital period with a period of $17.18\pm0.08$ years and an amplitude of $0.0018\pm0.0001$ d could be caused by the light-time effect (LITE) or magnetic activity of a dwarf star with a mass of 0.06 M$_\odot$, a third body in the system \citep{Zhu2021RAA....21...84Z}. \citet{Qian2001MNRAS.328..914Q}, using 35 photometric minimum times in detail and analysing the parabola-like change in \(O-C\) variations, which is a result of mass transfer, found that the decrease rate in the orbital period is dP / dt = - $4.39\times10^{-8}$ days/year, while \citet{Yang2003PASP..115..748Y} found it to be -$2\times10^{-8}$ days/year.  Yang also found an oscillation with a period of 16.1 days and an amplitude of $2.8\times10^{-7}$ days. However, in his subsequent study, he updated the period of this oscillation to $23.6\pm0.4$ and the amplitude to $A= 0.0028\pm0.0003$ days \citep{Yang2009AJ....137..236Y}. The cycle of these detected oscillations can be explained by magnetic activity or by the presence of a third object in the system \citep{Yang2009AJ....137..236Y}.

\begin{table}
\caption{Basic parameters of CC Com from the Simbad, 2MASS and TESS Catalogue.}
\begin{tabular}{llll}
\hline
Identifying information &   \\
\hline
TIC                       & 367683204  \\
2MASS ID                  & J12120602+2231586          \\
\textit{Gaia} ID          & DR3 4001489436181526656    \\
$\alpha_{2000}$           & 12:12:06.04    \\
$\delta_{2000}$           & 22:31:58.68                \\
\hline
Photometric  properties &   \\
\hline
G (Gaia)                  & 11$^{\rm m}$.19 (2)\\
B                         & 13$^{\rm m}$.09 (36)\\
V                         & 11$^{\rm m}$.42 (11)\\
2MASS J                   & 8$^{\rm m}$.986 (21)\\
2MASS H                   & 8$^{\rm m}$.341 (24)\\\hline
Stellar properties &   \\
\hline
Spectral type             & K4/5 V  \\
Period (CC Com A)      	  & 0.220\,d \\
Period (CC Com B)        & 7.9\,yr  \\
Period (CC Com C)        & 97.8\,yr  \\
$\pi_{\rm Gaia}$  (mas)   & 14.009(22) \\
\hline
\end{tabular}
\label{Table:cccom:basic:par}
\end{table}

In the present study, a comprehensive analysis of all available data sets from the literature, inclusive of recent observations and data from the Transiting Exoplanet Survey Satellite (TESS) in the Sectors 22 and 49 were conducted to accurately determine the orbital and physical parameters of the system, a quadruple system. Section 2 delineates the observations and the data reduction processes. Section 3 provides an in-depth analysis of the system's period change, examining all existing minima times in conjunction with newly acquired minima times. Furthermore, this section presents an analysis of all radial velocities of the very close system CC Com. The physical parameters of all constituent stars of the system are computed in Section~\ref{sec:Physical_par}. The final section encapsulates all the findings and discussions.

\section{OBSERVATIONS} \label{sec:obs}

\begin{figure}
\centering 
\includegraphics[scale=1.3]{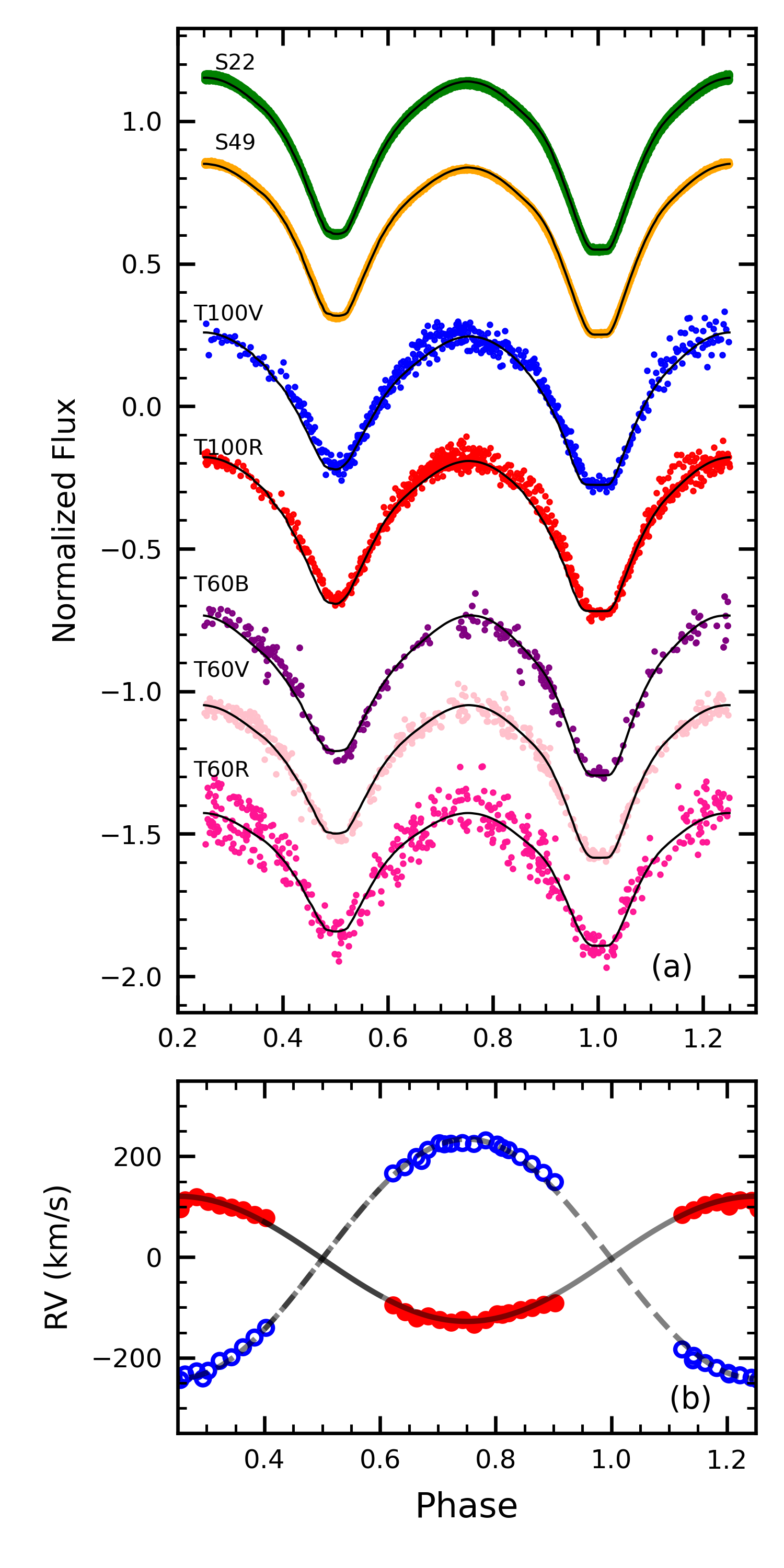}
\caption{Synthetic models (lines) of the system obtained during simultaneous solution with observed (coloured points) (a) light  and (b) radial velocity \citep{Pribulla2007AJ....133.1977P} curves. The light curves are randomly shifted along the y-axis for a better view.  See text for details.}
\label{Fig:cccom_all_lc_rv}
\end{figure}

New long-term multi-colour photometric monitoring of the system was performed on B, V and R filters using the iKon-L 936 BEX2-DD and the FLI ProLine 3041-UV CCD in the 0.6-meter TUG-T60 telescope between 2018 and 2021 years at the T\"UB\.ITAK National Observatory (TUG). Short-term optical observations of the system were made in V and R filters on 20 March, 3, 4 and 13 April 2018 with SI 1100 CCD using the 1-meter TUG-T100 telescope. The space-based sky survey programme TESS has sensitively observed CC Com and many other stars during different observing seasons. The system was observed in Sector 22 for 26.4 days, between 20 February and 17 March 2020, and in Sector 49 for 24.2 days, between 1 and 25 March 2020. 15998 data points were obtained in Sector 22, and 13512 in Sector 49. We downloaded the observations obtained with the TESS telescope from the MAST\footnote{https://mast.stsci.edu} servers. PDCSAP flux was used for the analyses and for the TESS data shown in Figure~\ref{Fig:cccom_all_lc_rv}. These data contain the simple aperture photometry (SAP) flux from which possible trends are extracted using cotrended basis vectors (CBVs) and are generally cleaner data than the SAP flux. The lighkurve \citep{Lightkurve_2018ascl.soft12013L} package was used to analyse and reduce the data sets of TESS observations.

The AstroImageJ \citep{Collins2017} programme was used to reduce round-based CCD observations. During the reduction process, flat, dark, and bias images obtained at observation nights were used. The magnitude of the system was then determined by the differential photometry method. The data processing and normalisation process was carried out with the same methods and programmes as we used for the \citet{Yakut2015MNRAS.453.2937Y, Cokluk2019MNRAS.488.4520C, Kocak2021ApJ...910..111K, Kocak2023PhDT.........3K} studies. Figure \ref{Fig:cccom_all_lc_rv} shows the CC Com light curves obtained from the TESS and TUG data sets. In this study, the expression given by Equation \ref{Eq:cccom:lineph} is used in the phase calculations of the system. Radial velocity observations of the system were obtained by \citet{Pribulla2007AJ....133.1977P}. The light curves obtained from the TESS observations of the system were solved simultaneously with the radial velocity curves. 
Figure 1 shows that the new multicolour observations obtained with ground-based telescopes are more scattered than the TESS observations. The precision of the observations obtained with ground-based telescopes is in the order of a few per cent, whereas that of TESS observations is in the order of a few thousandths. Although the observational sensitivity of space-based telescopes is much better, the lack of multi-colour observations can be seen as a deficiency. Multi-colour photometry from ground-based telescopes also has advantages in obtaining some parameters of binary systems (especially radiative parameters and colour variation for activity). 

\begin{equation} 
    \textrm{HJD~Min~I} = 24~39533\fd597(1)+0\fd2206868(7).
    \label{Eq:cccom:lineph}
\end{equation}

\begin{table}
\caption{Times of minima of the multiple system CC Com. All minima times are given as subtracted from 24~00000 days. The full table is given in the online version of this paper.} \label{table:mintimes:cccom}
\begin{tabular}{llll}
\hline
BJD         & Error     & Weight    &   Reference \\          
\hline
39533.5834	&   0.0002	& 	5	& 	1\\
42459.8869	&   0.0002	& 	5	& 	1\\
42461.8726	&   0.0002	& 	5	& 	1\\
42466.8380	& 	0.0002	& 	5	& 	1\\
42466.9485	& 	0.0002	& 	5	& 	1\\
42467.8313	& 	0.0002	& 	5	& 	1\\
42467.9411	& 	0.0002	& 	5	& 	1\\
42484.8239	& 	0.0002	& 	5	& 	1\\
42484.9335	& 	0.0002	& 	5	& 	1\\
42749.6458	& 	0.0002	& 	5	& 	1\\
42785.6194	& 	0.0002	& 	5	& 	1\\
42812.4327	& 	0.0020	& 	5	& 	1\\
\hline
\end{tabular}
\\
{References for Table~\ref{table:mintimes:cccom}. 
1-\citet{Zhu2021RAA....21...84Z}
2-\citet{Zejda2004IBVS.5583....1Z}; 
3-\citet{Safar2002IBVS.5263....1S}; 
4-http://var2.astro.cz/ocgate;
5-This study.
}
\end{table}

\section{DATA ANALYSIS} \label{sec:dataanalysis}
Using all available light curves and all published minima of the system, we have analysed the period change, light and radial velocity curves of the very close binary systems CC Com in individual sub-sections.

\subsection{Period Change Analysis}\label{sec:periodchange}

New observations of the system made with the ground-based TUG-T100, TUG-T60 and space-based TESS telescopes are shown in Figure \ref{Fig:cccom_all_lc_rv}. In this study, many times of minima have been calculated from the newly obtained observations. Table \ref{table:mintimes:cccom} lists some of the system's available times of minima. Four minima (at 58212.33157, 58212.44198, 58222.37265, and 58222.48422) were obtained from TUG T100 observations, and 120 minima from TESS observations. When reading minimum times from TESS observations, we calculated one minimum time by superimposing all three minima on top of each other. Therefore, we weighted the minimum times obtained from TESS by a factor of three. Other minimuma found in the literature were also collected, giving the system a total of 372. All times of minima were weighted according to the sensitivity of the observation and a period change analysis was performed with the MATLAB code developed by P. Zache \citep{Zasche2009NewA...14..121Z}. Equations \ref{Eq:cccom:3rdbody} and \ref{Eq:cccom:3rdbody2} were used in the solutions, under the assumption of sinusoidal and parabolic variation. While using Equations \ref{Eq:cccom:3rdbody} and \ref{Eq:cccom:3rdbody2} to obtain the parameters of the fourth system, we subtracted the long-term ($\sim$98-years) variation and applied Equation \ref{Eq:cccom:3rdbody2} to the remaining differences in Equation \ref{Eq:cccom:3rdbody}  for the fourth system and calculated the parameters of the possible short-term ($\sim$8-years) system with the help of Equation \ref{Eq:cccom:fm}.

\begin{figure}
\centering
\includegraphics[scale=0.58]{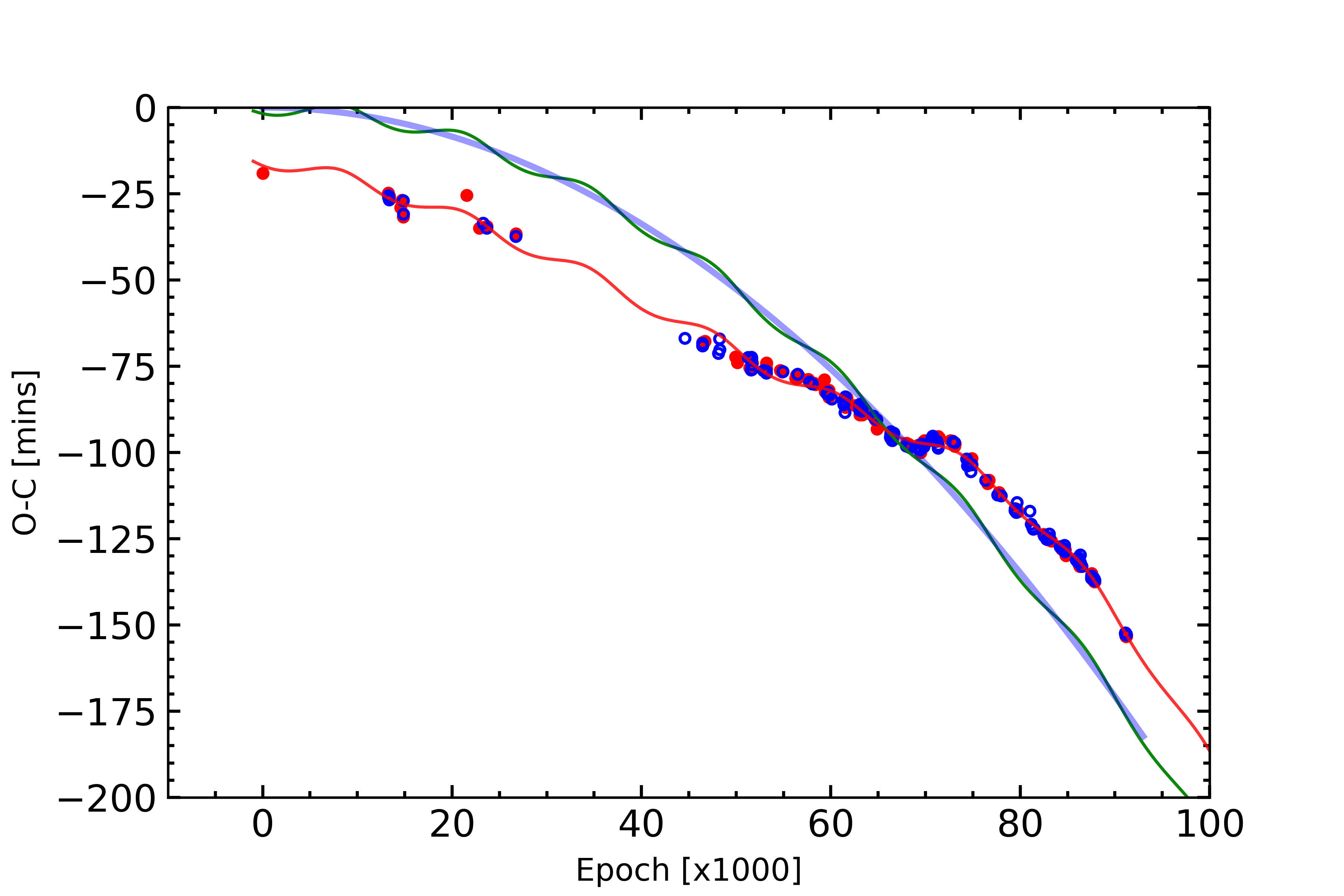}\\
\includegraphics[scale=0.58]{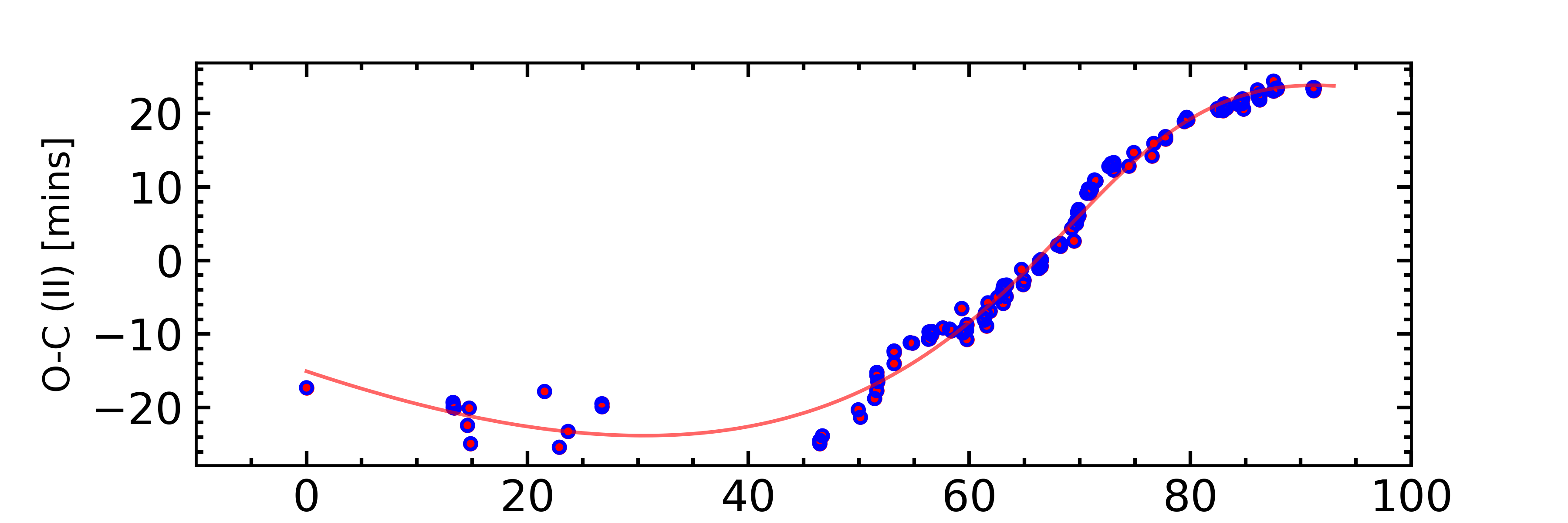}\\
\includegraphics[scale=0.58]{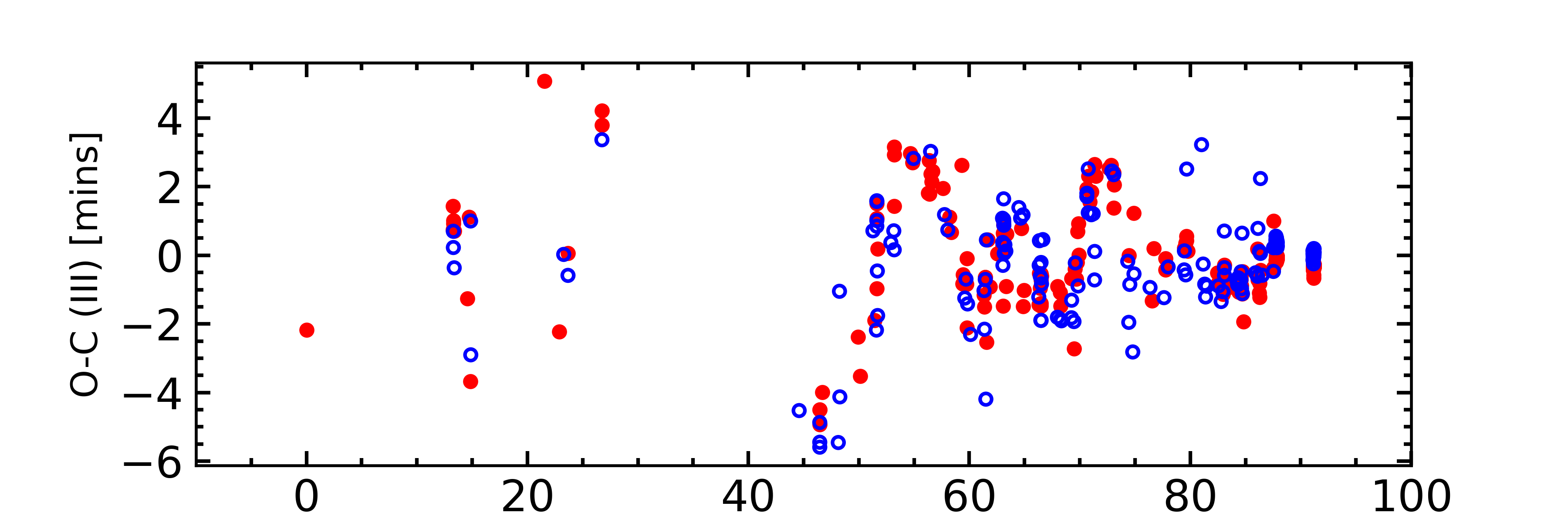}
\caption{Period change analysis of the multiple system CC Com. The top panel shows the observations and results of the model under different assumptions. The solid parabola represents the mass transfer between the components of CC Com A. The green solid sine-like variation represents the variation only in the presence of the third body with the parabolic variation. Finally, the solid line in red represents the variation with a third body with a period of 8 years and a fourth body with 98 years around CC Com~A. In the middle panel, we separately show the change in O-C (II) for the period of 98 years, and finally, in the bottom panel, we show the residuals.}
\label{Fig:cccom:oc}
\end{figure}

\begin{figure}
\centering
\includegraphics[scale=0.6]{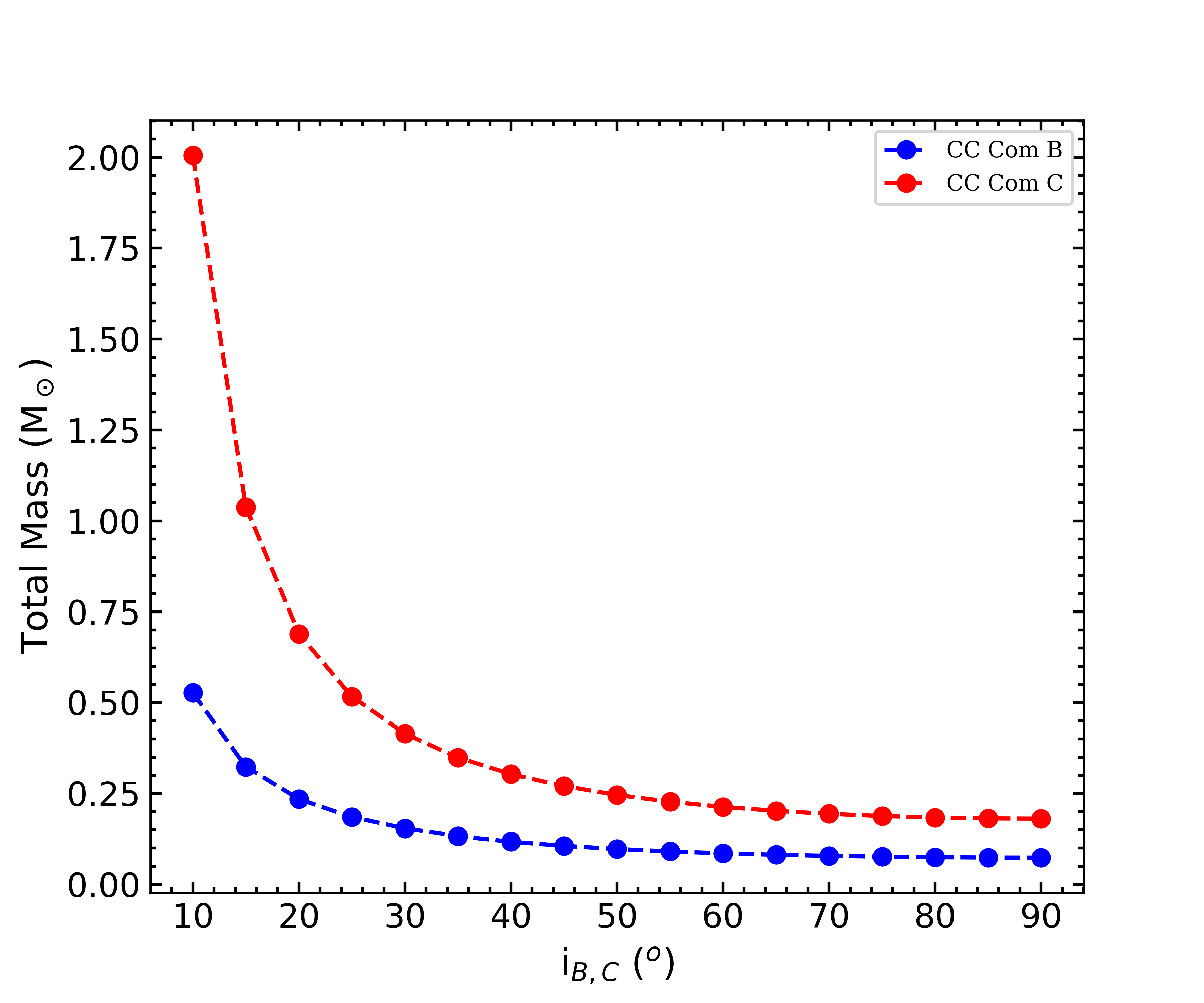}
\caption{Masses of the CC Com~B and CC Com~C components according to their orbital inclination angles.}
\label{Fig:cccom:ocmass34}
\end{figure}

Because CC Com is a contact binary system, and a parabolic variation is expected as a result of mass transfer between the components (Figure \ref{Fig:cccom:oc}, blue line). In addition to this parabolic variation, the presence of a third body is also taken into account (Equations \ref{Eq:cccom:3rdbody} and \ref{Eq:cccom:3rdbody2}). Analysis under the assumption of mass transfer and the presence of a third body reveals that a periodic change in the residuals remains. Therefore, a reanalysis was performed assuming a quadruple system and parabolic variation. The parameters obtained as a result of the period change analysis are given in Table \ref{table:cccom:OC}. Analyses have shown that there is a third object orbiting the binary system with a period of about eight years and that there may be a fourth object in the outer orbit with a period of about a century. The results obtained are shown in the O-C diagram in Figure \ref{Fig:cccom:oc}.
The masses of the third and fourth bodies orbiting CC Com A depend on the inclination of their orbital planes. The masses of the third and fourth bodies found in this study are shown in Figure~\ref{Fig:cccom:ocmass34} depending on the inclination of their orbital planes with respect to the plane of the sky.

The equations describing the minima are
\begin{equation}
\mathrm{Min\,I}=T_{0}+P_{0} E+\frac{1}{2} \frac{d P}{d E} E^{2} + \tau,  
\label{Eq:cccom:3rdbody}
\end{equation}
where
\begin{equation}
 \tau = \frac{a_{12} \sin i^{\prime}}{c}\times\left[\frac{1-e^{2}}{1+e^{\prime} \cos v^{\prime}} \sin \left(v^{\prime}+\omega^{\prime}\right)+e^{\prime} \sin \omega^{\prime}\right].
\label{Eq:cccom:3rdbody2}
\end{equation}

Mass functions are given by
\begin{equation}
f(m_3) = 4 \pi^2 \frac{\left(a_{12} \sin i\right)^3}{G P_{m}^2}=\frac{\left(M_3 \sin i\right)^3}{\left(M_1+M_2+M_3\right)^2},
\label{Eq:cccom:fm}
\end{equation}
f($m_3$) and f($m_4$) given in Table~\ref{table:cccom:OC} were calculated using Equation~\ref{Eq:cccom:fm}. In the above equations, the terms $i^{\prime}$, $e^{\prime}$, $v^{\prime}$ and $\omega^{\prime}$ denote the inclination of the orbit, the eccentricity, the true anomaly and the longitude of the periastron from the ascending node of the third component of the system, respectively.

\begin{table}
\caption{Results of the $O-C$ analysis of the binary system CC Com.} \label{table:cccom:OC}
\begin{tabular}{ll}
\hline
Parameter                                      & Value                             \\
\hline $T_o$ (${\rm JD}/{\rm{d} - 2400000}$)   & 39533.597$\pm 0.0016$   \\
$P_o/d$                                        & 0.22068684$\pm 0.00000002$        \\
&\\
$P_3/d$                                        & 2893$\pm 29$     \\
$P_3/yr$                                       & 7.92$\pm 0.08$           \\
$T_3$   (${\rm JD}/{\rm{d} - 2400000}$)        & 2427095$\pm 28$ \\
$e_3$                                          & 0.065        \\
$\omega_3/^\circ$                              & 122$\pm 9$   \\
$A_3/d$                                        & $0.00153\pm 0.00008$     \\
$f(m_3)/{\rm{M_\odot}}$                        & 0.000300$\pm 0.000005$       \\
$m_{3;i'=30^\circ}/{\rm{M_\odot}}$             & 0.16        \\
$m_{3;i'=60^\circ}/{\rm{M_\odot}}$             & 0.086        \\
$m_{3;i'=90^\circ}/{\rm{M_\odot}}$             & 0.074       \\
$P_4/d$                                        & 35720$\pm 2147$     \\
$P_4/yr$                                       & 97.8$\pm 5.8$                         \\
$T_4$  (${\rm JD}/{\rm{d} - 2400000}$)         & 2455657$\pm 342$               \\
$e_4$                                          & 0.439 $\pm 0.034$               \\
$\omega_4/^\circ$                              & 25$\pm 6$                       \\
$A_4/d$                                        & $0.0165\pm 0.0002$     \\
$f(m_4)/{\rm{M_\odot}}$                        & 0.00318$\pm 0.00001$           \\
$m_{4;i'=30^\circ}/{\rm{M_\odot}}$             & 0.41                              \\
$m_{4;i'=60^\circ}/{\rm{M_\odot}}$             & 0.21                              \\
$m_{4;i'=90^\circ}/{\rm{M_\odot}}$             & 0.18                              \\
\hline
\end{tabular}
\end{table}

Using the Period04, we performed a Fourier analysis on the residuals of the O-C analysis (Figure 3a) and looked for possible variations. In this analysis, we found a dominant frequency at 0.00061(1) c/d.  This corresponds to a variation of about 8.9$\mp0.2$ years, which is similar to the value obtained from our period analysis under the fourth body assumption.  Of course, the distribution of a limited number of minimum times distributed over 55 years prevents us from obtaining this value more precisely.

\subsection{Modelling of Light and Radial Velocity Curves }

The simultaneous solution of the CC Com spectral data and precise observations from the TESS satellite with the Wilson-Devinney \citep{Wilson1971ApJ...166..605W, Wilson1979ApJ...234.1054W} and Phoebe \citep{prsa2005ApJ...628..426P} programmes allowed us to accurately determine the physical and orbital parameters of the components.

Although there are many light curves of the system obtained with ground-based telescopes, the light curves recently obtained with TESS observations are very sensitive. Multi-colour BVR photometric observations with ground-based telescopes are at least as important, if not more precise, than the TESS Sectors 22 and 49. While fitting the synthetic models, we first solved the TESS (S49) data sets simultaneously with the existing radial velocity data sets \citep{Pribulla2007AJ....133.1977P,McLean1983MNRAS.203....1M,Rucinski1977PASP...89..684R} in order to obtain more precise orbital parameters. The solutions show that the cooler component has a larger radius and a higher mass than the hotter star. The result of the solution is shown in Figure \ref{Fig:cccom_all_lc_rv} as a solid line over the observations. The mean temperature of the hot star (T$_{\rm A1}$=4300~K), gravitational darkening coefficients (g$_{\rm A1}$=g$_{\rm A2}$) \citep{Lucy1967ZA.....65...89L}, the albedos (A$_{\rm A1}$=A$_{\rm A2}$) \citep{Rucinski1969AcA....19..245R} and the logarithmic limb darkening coefficients (g$_{\rm A1}$=g$_{\rm A2}$) \citep{Claret2018A&A...618A..20C} are taken as fixed parameters during the analysis of the light curve. The orbital inclination ($i$), the semimajor axis of the relative orbit $a$, the radial velocity of the binary centre of mass (V$_\gamma$), the potential of the cold components ($\Omega _{\rm A1=A2}$), the temperature of the secondary component (T$_{\rm A2}$), the luminosities and the third light contribution (l$_{\rm B+C}$), the spot parameters and the mass ratio $q$ were adjustable parameters. 

After several model runs, we stopped modelling when the correction of each parameter was well below the errors. We presented the parameters obtained with their errors in Table \ref{tab:cccom:lcrv} and the comparison of the synthetic model with the observations in Figure \ref{Fig:cccom_all_lc_rv}. As can be seen, the model and the observations are in good agreement. Preliminary analyses were performed using the combined solution of Sector 49 and radial velocities. To represent the asymmetries seen at the maximum phases of the light curves, it was necessary to use spotted models during the fitting. For this reason, three spots, whose properties are given in Table \ref{tab:cccom:lcrv}, were taken into account during the solution. When solving for the observations in Figure \ref{Fig:cccom_all_lc_rv} (Sector 22 and other ground-based observations), the fundamental parameters (P, i, q, T$_2$, etc.) were fixed and only the limb darkening, relative luminosity, and spot parameters were allowed as free parameters. In this way, it is clear from the solid lines in Figure \ref{Fig:cccom_all_lc_rv} that the parameters given in Table \ref{tab:cccom:lcrv} are compatible with all observations.

\begin{table}
\caption{The results of light curves analysis with their formal
1$\sigma$ errors for CC Com. See text for details. This solution can be represented by the three spots (S1, S2, S3) on the surface of the primary star.}\label{tab:cccom:lcrv}
\begin{tabular}{ll}
\hline
Parameters& Results\\
\hline
$T_{0}$ (${\rm JD}/{\rm{d} - 2400000}$)   		& 58900.09377(8) \\
$P/{\rm d}$                               		& 0.220684(1)   \\
Orbital inclination, $i$ ($^\circ$)             & 89.9(1)	 \\
$\Omega _{\rm 1}$  = $\Omega _{\rm 2}$                             & 5.11(1)	  \\
$q = m_{\rm 1}/m_{\rm 2}$                       & 1.913(23)   \\
Filing factor, f(\%)                             & 7     \\
$T_{\rm 1}/{\rm K}$                             & 3930 (45)   \\
$T_{\rm 2}/{\rm K}$                             & 4300      \\

Fractional radius of secondary($R_{\rm 1}/a$)  & 0.5392(3)     \\
Fractional radius of primary ($R_{\rm 2}/a$)    & 0.4307(3)     \\

Luminosity ratios:                         		& 				 \\
$l_{\rm 1}/l_\textrm{total}$                    & \%57         \\
$l_{\rm 2}/l_\textrm{total}$                    & \%43         \\
Spot (S) parameters:    & (S1, S2, S3)      \\
Colatitude  ($^{\circ}$)        &  (90, 90, 90)    \\
Longitude   ($^{\circ}$)        &  (140, 30, 330)     \\
Radius      ($^{\circ}$)        &  (17, 15, 15)     \\
Temperature factor (T$_{\rm spot}$/T$_{\rm eff}$) &  (0.9, 0.9, 0.9)     \\
\hline
$K_{\rm 1}/{\rm km\,s^{-1}}$                    & 237.7(3.6)       \\
$K_{\rm 2}/{\rm km\,s^{-1}}$                    & 124.2(1.9)       \\
$V_\gamma/{\rm km\,s^{-1}}$  	& $3.4$(7)       \\
a$_{\rm 1}$sin$i/R_{_\odot}$                    & 0.542(3)     \\
a$_{\rm 2}$sin$i/R_{_\odot}$                    & 1.036(5)     \\
m$_{\rm 1}$sin$^3i/M_{_\odot}$                  & 0.712(9)       \\
m$_{\rm 2}$sin$^3i/M_{_\odot}$                  & 0.372(5)       \\
\hline
\end{tabular}
\end{table}

\begin{table}
\begin{center}
\caption{Astrophysical Parameters of CC Com A. The standard errors $\sigma$ are given in parentheses in the last digit quoted.}
\label{table:cccom:PhyPar}
\begin{tabular}{lll}
\hline
Parameter                                     & A1  		  &  A2\\
\hline
Mass $M/\rm{M_\odot}$                   & $0.712$(9)     	  & $0.372$(5)    \\
Radius $R/\rm{R_\odot}$                 & $0.693$(6)      	  & $0.514$(5)     \\
Temperature $T_{\rm eff}/{\rm K}$       & $3\,930$(45)    	  &  $4\,300(55)$  \\
Luminosity  $L/\rm{L_\odot}$            & $0.103$             & $0.081$     \\
Surface gravity $\log_{10}(g/\rm{cm\,s^{-2}})$ & $4.609$    & $4.587$        \\
Bolometric magnitude $M_B$              & $7^{\rm m}.20$      	  & $7^{\rm m}.46$      \\
Absolute magnitude  $M_V$               & $8^{\rm m}.25$     	  & $8^{\rm m}.21$      \\
Distance between components $a/\rm{R_{_\odot}}$  &~~~~~~~~~~~~~~~~~~$1.578(8)$ &		\\
Orbital period change $\dot{P}$ ($day yr^{-1}$) &~~~~~~~~~~~~~~~~$-4.8\times10^{-8}$   &    	\\
Mass transfer rate  $\dot{M}$ ($\rm{M_\odot} yr^{-1}$)&~~~~~~~~~~~~~~~~$-5.7\times10^{-8}$  & 	\\
Distance  $d/{\rm parsec}$                            &~~~~~~~~~~~~~~~~~~~~~~$69(4)$    &      	\\
\hline
\end{tabular}
\end{center}
\end{table}

\section{PHYSICAL PARAMETERS OF THE SYSTEM}\label{sec:Physical_par}

Previous studies of CC Com have mentioned the possibility of mass transfer between its components and the existence of a third body gravitationally bound to the system. Period change analysis revealed that CC Com is actually a possible quadruple system for the first time in this study. The physical parameters of the component stars, derived from synthetic models using CC Com's high-precision TESS observations and radial velocity data, are the most accurate ever obtained. Figure \ref{Fig:cccom_all_lc_rv} shows the solved TESS light curves and the fitted synthetic models, while Table \ref{table:cccom:PhyPar} shows the parameters obtained from the solution.

Double-line eclipsing binary stars are still the most reliable for determining the accurate physical parameters of companion stars. It is desirable to model double-lined radial velocity curves simultaneously and, if possible, a large number of light curves to determine the various parameters of the components of a binary star, such as mass, radius, and luminosity well. In this study, CC Com's radial velocity curves and several light curves were analysed simultaneously. These quantities obtained as a result of the solution, with their uncertainties, are given in Table \ref{tab:cccom:lcrv} and Figure \ref{Fig:cccom_all_lc_rv}.

The calculations took the Sun's effective temperature to be 5777 K and its bolometric magnitude of 4.775 mag. As a result of our analyses, we obtained the mass of the cooler component as M$_{\rm  A1}$ = 0.712 M$_\odot$, the mass of the hotter component as M$_{\rm  A2}$ = 0.372 M$_\odot$ and the radii as R$_{\rm  A1}$ = 0.693 R$_\odot$ and R$_{\rm  A2}$ = 0.514 R$_\odot$, respectively. The distance of the system we obtained in this study was 69 pc. The results showed that the hot component has a smaller radius, mass, and luminosity. The visual magnitudes of the components were calculated with the results of the tables of \citet{Pecaut2013ApJS..208....9P} for the bolometric corrections. The total masses of CC Com B and C depend on the orbital inclination angles and are shown in Figure \ref{Fig:cccom:ocmass34} in blue and red colours, respectively. Accordingly, the masses that the component stars can have at $i=90^\circ$ are 0.074 for CC Com B and 0.18 for CC Com C.

\section{DISCUSSION AND CONCLUSION}\label{sec:results}

By studying binary and multiple stars photometrically and spectroscopically, we can accurately determine their basic physical and orbital parameters, such as period, luminosity, temperature, etc. The parameters obtained, such as mass, radius, temperature, luminosity, and absolute magnitude, provide very important information about their formation, evolution, and end of life. The observed binary systems exhibit considerable diversity in their mass and period distributions. Depending on their initial conditions (mass, chemical abundance, orbital period, etc.), the binary systems' orbits and the component stars' evolution can be fast or slow. In the case of close binaries, such as CC Com, the evolution becomes much more complicated \citep{Yakut2005ApJ...629.1055Y,Icli2013AJ....145..127I}. Close binary systems are important astrophysical objects because they allow us to test critical physical phenomena such as angular momentum loss, mass transfer, and mass loss. 

The very short orbital periods and extremely low mass ratios of the component stars are particularly important from an evolutionary point of view. Examples of binary systems with very small mass ratios are SX Crv \citep{Rucinski2006AJ....132.1539R}, V870 Ara \citep{Poro2021OAst...30...37P}, KR Com \citep{2010A&A...519A..78Z} and V1191 Cyg \citep{2012NewA...17...46U}. Only seven systems with very short orbital periods (P<0.25 d) have been studied both photometrically and spectroscopically, and their parameters are well known. CC Com is one of these systems and the others are SDSS J001641-000925 (0.199 d, \citet{Davenport2013ApJ...764...62D}), OT Cnc (0.218 d, \citet{2021MNRAS.501.2897G}), J160156 (0.227 d, \citet{Lohr2014A&A...563A..34L}), J093010 B (0.228 d, \citet{2021MNRAS.501.2897G}), V523 Cas (0.234 d, \citet{Kose2009Ap&SS.323...75K}) and RW Com (0.237 d, \citet{ozavci2020AcA....70...33O}). Such very short-period binary systems are excellent laboratories for understanding the nature of interacting binary stars and studying merger processes, mass transfer and mass loss.

In this study, the nature of CC Com has been analysed in detail. In addition to the new observations obtained with the TUG telescopes, all available light curves of the system observed by TESS were collected for use in the light curve analysis. As a result of the simultaneous analysis of the light and radial velocity curves, the orbital and physical parameters of the component stars are given in Tables~\ref{tab:cccom:lcrv} and \ref{table:cccom:PhyPar}. It was found that the cooler component has a larger mass and radius than the hotter component. We combined 221 minima from TUG and TESS observations with those found in the literature. We then analysed them for period change. As a result of our analysis, it was determined, for the first time, in this study that the system is a possible quadruple system (A1+A2, B, C). CC Com A is a binary system with a period of 0.220 days. The period of the middle orbiting star (B) is 7.9 years and the period of the outer orbiting star (C) is 98 years. In this study, using the distance modulus method, the distance of CC Com was found to be 69 pc, which is in agreement with Gaia \citep{gaia2021A&A...649A...1G} by the astrometric method (71.4 pc).

The activity seen in late-type stars and the presence of spot regions on the stellar surface because of this activity can cause distortions in the light curves of the binary system. If the solutions are assumed to be spotless, the parameters of the binary system may be incorrectly estimated. Therefore, the light curves of active binary systems should be solved under spotted model assumptions. As seen in Figure~\ref{Fig:cccom_all_lc_rv}, a slight difference in level is observed in the maximum phases of the light curves of CC Com. This level difference is the O'Connell effect and is caused by spots on the stellar surface. During the solution of the light curves, we first applied a synthetic model to the observations of TESS, which gives the most sensitive light curve, and obtained the possible spots. Then, we fixed the system parameters we obtained and performed solutions by adjusting different spot parameters (spot latitude, spot longitude, spot radius and a temperature factor), limb darkening, and relative luminosity as free parameters. In this way, we also modelled the TUG T100 and T60 data sets. As seen in the figures, all observations and models are in good agreement.

\begin{acknowledgement}
The author sincerely thanks K. Yakut and C. Tout for their careful review of the manuscript and insightful recommendations and thank to anonymous referee for comments and helpful constructive suggestions, which helped us improve the paper. This study was supported by the Scientific and Technological Research Council of T\"urkiye (T\"UB\.ITAK Prj 122F474 and 117F188). DK thanks T\"UB\.ITAK-2219 for her scholarship. The author thanks the numerous people who have helped make the T\"UB\.ITAK National Observatory (Prj no:18AT60-1301) and the NASA {\it{TESS}} mission possible.
\end{acknowledgement}

\paragraph{Funding Statement}
This study was supported by the Scientific and Technological Research Council of Turkey (T\"UB\.ITAK Prj 2219).

\paragraph{Competing Interests}
None

\paragraph{Data Availability Statement}
The data used in this study are given as online tables of the TUG data sets. In addition, TESS satellite data was used in some of the analyses and can be obtained from the MAST data archive at {https://mast.stsci.edu}. 

\printendnotes

\printbibliography

\end{document}